# Anisometric Charge Dependent Swelling

# of Porous Carbon in an Ionic Liquid

F. Kaasik,[1,2] T. Tamm,[2] M. M. Hantel,[3] E. Perre,[1]

A. Aabloo,[2] E. Lust,[4] M. Z. Bazant,[5,6] V. Presser[1,7*]

[1] *INM – Leibniz Institute for New Materials, Energy Materials Group, 66123 Saarbrücken, Germany*

[2] *IMS Lab., Institute of Technology, University of Tartu, 50411 Tartu, Estonia*

[3] *General Energy Research Department, Paul Scherrer Institut, CH-5232 Villigen PSI, Switzerland*

[4] *Institute of Chemistry*, University of Tartu, 50411 Tartu, Estonia

[5] *Department of Chemical Engineering, Massachusetts Institute of Technology, Cambridge, MA 02139, United States*

[6] *Department of Mathematics, Massachusetts Institute of Technology, Cambridge, MA 02139, United States*
[7] *Department of Materials Science and Engineering, Saarland University, 6613 Saarbrücken, Germany*

[*] *Corresponding author's email: volker.presser@inm-gmbh.de*

**Abstract**

*In situ* electrochemical dilatometry was used to study, for the first time, the expansion behavior of a porous carbon electrode in a pure ionic liquid, 1-ethyl-3-methyl-imidazolium-tetrafluoroborate. For a single electrode, an applied potential of -2 V and +2 V against the potential of zero charge resulted in maximum strain of 1.8 % and 0.5 %, respectively. During cyclic voltammetry, the characteristic expansion behavior strongly depends on the scan rate, with increased scan rates leading to a decrease of the expansion. Chronoamperometry was used to determine the equilibrium specific capacitance and expansion. The obtained strain *versus* accumulated charge relationship can be fitted with a simple quadratic function. Cathodic and anodic expansion data collapses on one parabola when normalizing the surface charge by the ratio of ion volume and average pore size. There is also a transient spike in the height change when polarity is switched from positive to negative that is not observed when changing the potential from negative to positive indicating the size and the shape of the ion is influencing the expansion behavior.

**Keywords**

Electrochemical dilatometry; ionic liquid; electrical double layer capacitor; porous amorphous carbon; carbide-derived carbon

## 1. Introduction

Electrical double-layer capacitors (EDLC), more commonly known as ultra- or supercapacitors, have attracted significant interest for energy storage and management because of their high power density, high charge/discharge efficiency, and long service life.[1] Recently, it has been demonstrated that the process of charge storage *via* ion electrosorption is associated with volume variations of porous carbon electrodes upon successive charge/discharge cycles [2-4]. The magnitude of height expansion varies significantly depending on the carbon material used and three different kinds of electrode expansion have been identified so far:

- (i) Expansion due to initial electrode conditioning. This process corresponds, for example, to carbon nanotubes de-bundling (≈10 % irreversible expansion) [5].
- (ii) Expansion in the regime of electrochemical stability yielding fully reversible expansion in the range of *circa* 1 % [5, 6].
- (iii) Expansion related to intercalation, such as intercalation of acetonitrile-solvated/desolvated tetrafluoroborate ions into graphitic carbon causing strain of up to 60 % [3].

Until now, two *in situ* methods have been employed to study the expansion behavior of supercapacitor electrodes: electrochemical dilatometry (eDilatometry) in organic electrolytes [2-7] and electrochemical atomic force microscopy (eAFM) using an ionic liquid with added lithium salt [8]. As shown in these studies, carbon expansion is an intrinsic property that occurs with or without the presence of a polymer binder. The expansion also depends on the ion size, with enhanced swelling occurring when the average pore size is below the dimensions of the ions [6]. The observed level of swelling differs for positive and negative polarization, the latter usually exhibiting higher electrode expansion because of the large size of the cations [7].

In this paper, for the first time, the swelling of electrodes is reported in a pure ionic liquid (IL; 1-ethyl-3-methylimidazolium tetrafluoroborate, $[Emim]^+[BF_4]^-$). Our work was motivated by the increasing interest in ILs because of their low vapor pressure, non-flammable character, and wide electrochemical stability window. The study of the expansion behavior of carbon in ILs is equally important for supercapacitors and ionic electroactive polymer composite actuators [9], as both technologies rely on the formation kinetics of the electrical double-layer.

## 2. Experimental

*2.1 Material characterization and electrode preparation*

Carbide-derived carbon (CDC)[10] obtained by chlorine treatment of boron carbide ($B_4C$) was purchased from Y-Carbon Inc. (USA) and commercial grade YP50 activated carbon was obtained from Kuraray Co. Ltd (Japan). The CDC material was characterized with $N_2$ gas sorption at -196.15 °C using a Quantachrome (USA) Autosorb-6 system; the pore characteristics were calculated using the Brunauer-Emmett-Teller Equation (BET) [11] and quenched solid density functional theory (QSDFT) [12].

Electrodes were prepared by mixing the porous carbon $B_4C$-CDC for the working electrode (WE; thickness: 90 µm, diameter: 8 mm, mass: 3.0±0.1 mg) and activated carbon for the counter electrode (CE; thickness: 110 µm, diameter: 18 mm, mass: 17.8±0.1 mg); both mixtures contained 5 mass% polytetrafluorethylene (60 mass% dispersion in water from Sigma-Aldrich) and were roll-pressed to form freestanding films.

*2.2 eDilatometry setup and electrochemical measurements*

An ECD-nano-DL eDilatometer (similar to the setup in Ref. [3]) from EL-Cell (Germany) was used with gold current collectors. The electrochemical cell was assembled in an Argon filled glovebox ($O_2$, $H_2O$ < 1 ppm) and $[Emim]^+$ $[BF_4]^-$ (purity ≥99.0%, Sigma Aldrich) was used as the electrolyte. Electrochemical and dilatometric measurements were performed at 20±1 °C with a Biologic VSP-300 potentiostat and displacements were recorded with an Agilent 34972 A data acquisition system. The expansion, perpendicular to the separator, measured with an accuracy of approximately 5 nm, was normalized by the electrode thickness and the point of zero strain was corrected for instrumental shifts. To facilitate data comparison, the net zero expansion at 0 V was normalized to its initial value after each cycle.

**3. Results and discussion**

*3.1 Porosity analysis of the electrode material*

$B_4C$-CDC shows a specific surface area $S_{BET}$ of 1405 m$^2$/g ($S_{QSDFT}$ = 1375 m$^2$/g) in agreement with literature [13, 14]. This material exhibits a bimodal pore size distribution with maxima at ≈8 and ≈17 Å and a volume-weighted average pore size ($d_{50}$) of 16 Å. The porosity is 0.92 cm$^3$/g with ≈60 vol% of the pore volume associated with micropores (< 20 Å) and ≈20 vol% associated with ultramicropores (< 8 Å). $B_4C$-CDC with these pore characteristics was chosen considering the ionic radii (values from DFT calculations) [15] of $[Emim]^+$ (approx. 9.5 Å x 6.3 Å x 5.3 Å; $V_{ion}$ = 118.8 Å$^3$) and $[BF_4]^-$ (approx. 4.7 Å x 4.9 Å x 4.7 Å; $V_{ion}$ = 48.6 Å$^3$) to enable access to the majority of pores.

*3.2 eDilatometry and electrochemical characterization*

The cyclic voltammograms (CVs) and the corresponding expansions are depicted in **Fig. 1a-c** in the case of potential sweeping between +2 V and -2 V. Starting from a nearly rectangular-shaped CV typical for EDLCs at 5 mV/s, the influence of resistance (IR-drop) becomes more pronounced as the scan rate is increased (**Fig. 1a**). All recorded strain was reversible, with larger expansions observed at negative polarization. The smaller accumulated charge at higher scan rates is accompanied by a smaller expansion and flatter hysteresis and the difference between the maximum anodic and cathodic expansion decreases with increasing scan rate (**Fig. 1b**). This can be explained by the limited ion migration speed of the ionic liquid inside porous carbon at room temperature. Additionally, the increasing scan rate leaves too little time to reach the adsorption equilibrium.

The data at 5 mV/s shows that the expansion continues even after reversal of the potential scan direction (labeled $\varepsilon_e^-$ and $\varepsilon_e^+$ in **Fig. 1c**) and continued expansion can be observed until the net current is zero. In the literature, this dynamic behavior has also been observed for microporous CDC[6] and activated carbon in 1M tetraethylammonium tetrafluoroborate in acetonitrile [5]; yet, it is much more pronounced in our study as a result of the lower ion mobility in ILs [16].

Data from chronoamperometric charge- and discharge time (*t*) for 1 h was used to calculate the equilibrium specific capacitance and the corresponding expansion at different potentials. As shown in **Fig. 1d**, the capacitance is increasing with the potential and reaches 96 F/g and 98 F/g at -2 V and +2 V, respectively, compared to 70 and 78 F/g at -0.5 V and +0.5 V, respectively. In qualitative agreement with the CV results, the expansion during the cathodic polarization is larger than during the anodic polarization; however, the equilibrium expansion (meaning expansion at $t \rightarrow \infty$) is up to ≈200% larger than the maximum dynamic expansion during CV cycling at 5 mV/s.

Charging of $B_4C$-CDC electrode to +2 V or -2 V with subsequent discharging to 0 V yields distinct expansion minima and maxima (**Fig. 2a**). When the polarity is reversed from -2 V to +2 V or *vice versa*, the system polarizes too rapidly inhibiting the system from achieving the same expansion experienced when fully discharged. Yet, there is still an expansion minimum between the cathodic and anodic expansion associated with the migration of formerly electrosorbed ions into the electrolyte reservoirs found between the carbon particles and inside the larger pores. Interestingly, the expansion behavior at the moment of potential inversion from +2 V to -2 V or from +2 V to 0 V yields a small expansion spike before the swelling recedes **(Fig.2b).** The spike is smaller in amplitude when choosing +1 V instead of +2 V and it does <u>not</u> occur if potential is changed from -2 V to either +2 V or 0 V, that is, when cations are desorbed and anions are adsorbed. The reason for this short-lived phenomenon remains unclear, but may be related to compression, electrostatic repulsion, and re-orientation which is different for the spherical $BF_4^-$ and the flat anisometric $[Emim]^+$ [15].

Fitting of the strain ($\varepsilon$) as a function of charge ($Q$) has not been addressed in the literature before. We see that a quadratic equation ($y=A \cdot x^2$) is sufficient to describe the experimental data with a larger value for $A$ in case of cation insertion ($A_{cation} / A_{anion}$ = 4.1). The correlation $\varepsilon \propto |Q|^2$ is similar to the relationship between the repulsive pressure of overlapping double-layers and the square of the applied charge $\left(P \propto |Q|^2\right)$ in the low-voltage Debye-Hückel approximation [17, 18], but the theory must be modified for ion crowding and overscreening in ionic liquids [19, 20]. At high voltage, condensed layers of counterions form at the maximum density $V^{-1}_{ion}$, and large steric repulsion occurs when their total width, scaling as $|Q| \cdot V_{ion}$ [21, 22] becomes comparable to the pore size. Indeed, as shown in **Fig. 2c**, the strain caused by cation and anion insertion collapses onto one parabola when plotted *versus* the normalized charge $Q_{norm} = |Q|/(V^{-1}_{ion} \cdot S_{BET} \cdot m \cdot d_{50})$ scaled to the product of the maximum counterion density, specific surface area, electrode carbon mass, and average pore size, respectively.

At the electrode potential of -2 V, we observed an expansion equivalent to 5 vol%, assuming purely isometric expansion. However, when we extrapolate the data to -3.5 V, the strain becomes 12 % (≈*40 vol%* expansion) while the positively charged electrode at 3.5 V would show only 3 % (≈10 *vol%* expansion) increase. In this case, the completed cell would discharge up to 7 V (a value theoretically possible for ideal conditions) and electrode strain may become significant to consider as a technological design parameter, whose design principles will differ from ionomer based systems.[21]

## 4. Conclusions

This dilatometric study of porous carbide-derived carbon in pure [Emim]$^+$[BF$_4$]$^-$ shows reversible expansion, depending on the applied potential. Moreover, the expansion depends not only on the applied charge, but also on the ion size. Our data on microporous carbon implies that the steric repulsion between condensed counterion layers in solid nanopores plays a major role since the anodic and cathodic electrode expansion can be fitted with one quadratic equation when the surface charge is corrected for the different ion sizes and the average pore size. The measured expansion of carbon in IL at a half cell potential of -2 V corresponds to a volumetric expansion of around 5 vol%. With the research activities related to further increasing the electrochemical stability and operation potential, the electrode expansion may ultimately become a limiting factor for the EDLC performance, including cycle life. The high expansion levels for EDLC materials at higher potentials also hold promise for high-performance actuators and strain sensors based on porous carbon in ionic liquids whose design principles will differ from ionomer based systems [21].

**Acknowledgements**

FK acknowledges funding from a Johann von Seidlitz Fellowship from University of Tartu Foundation funded by Mr. Jürgen Lambrecht and from the European Social Fund's (ESF) Doctorial Studies and Internationalisation Programme DoRa carried out by Archimedes Foundation. FK acknowledges support from the graduate school "Functional materials and processes" which receives funding from the ESF under project 1.2.0401.09-0079. VP and EP acknowledge funding from the German Federal Ministry for Research and Education (BMBF) in support of the nanoEES$^{3D}$ project (award number 03EK3013) as part of the strategic funding initiative energy storage framework. VP thanks Prof. Eduard Arzt for his continuing support. The authors thank Dr. Mesut Aslan for help with gas sorption measurements and Dr. Jennifer S. Atchison for helpful discussions (both at INM).

# References

[1] F. Beguin, E. Frackowiak, Supercapacitors, Wiley, Weinheim, 2013.
[2] M. Hahn, O. Barbieri, F.P. Campana, R. Kötz, R. Gallay, Appl. Phys. A: Mater. Sci. Process. 82 (2006) 633.
[3] M. Hahn, O. Barbieri, R. Gallay, R. Kötz, Carbon 44 (2006) 2523.
[4] P.W. Ruch, M. Hahn, D. Cericola, A. Menzel, R. Kötz, A. Wokaun, Carbon 48 (2010) 1880.
[5] P.W. Ruch, R. Kötz, A. Wokaun, Electrochim. Acta 54 (2009) 4451.
[6] M.M. Hantel, V. Presser, R. Kötz, Y. Gogotsi, Electrochem. Commun. 13 (2011) 1221
[7] M.M. Hantel, V. Presser, J.K. McDonough, G. Feng, P.T. T. Cummings, Y. Gogotsi, R. Kötz, J. Electrochem. Soc. 159 (2012) A1897
[8] T.M. Arruda, M. Heon, V. Presser, P.C. Hillesheim, S. Dai, Y. Gogotsi, S.V. Kalinin, N. Balke, Energy Environ. Sci. 6 (2013) 225
[9] J. Torop, M. Arulepp, J. Leis, A. Punning, U. Johanson, V. Palmre, A. Aabloo, Materials 3 (2009) 9.
[10] V. Presser, M. Heon, Y. Gogotsi, Adv. Funct. Mater. 21 (2011) 810.
[11] S. Brunauer, P.H. Emmett, E. Teller, J. Am. Chem. Soc. 60 (1938) 309
[12] A.V. Neimark, Y. Lin, P.I. Ravikovitch, M. Thommes, Carbon 47 (2009) 1617
[13] H. Wang, Q. Gao, Carbon 47 (2009) 820.
[14] A. Jänes, L. Permann, M. Arulepp, E. Lust, Electrochem. Commun. 6 (2004) 313
[15] E. Soolo, D. Brandell, A. Liivat, H. Kasemägi, T. Tamm, A. Aabloo, J. Mol. Model. 18 (2012) 1541
[16] O. Borodin, J. Phys. Chem. B 113 (2009) 11463.
[17] P.M. Biesheuvel, J. Colloid Interface Sci. 238 (2001) 362.
[18] J. Israelachvili, Intermolecular and Surface Forces: With Applications to Colloidal and Biological Systems., Academic Press, Amsterdam, 1991.
[19] M.Z. Bazant, B.D. Storey, A.A. Kornyshev, Physical Review Letters 106 (2011) 046102.
[20] A.A. Kornyshev, The Journal of Physical Chemistry B 111 (2007) 5545
[21] A.L. Alpha, H.C. Ralph, A.K. Alexei, Journal of Physics: Condensed Matter 25 (2013) 082203.
[22] M.S. Kilic, M.Z. Bazant, A. Ajdari, Physical Review E 75 (2007) 021502.

**Figures**

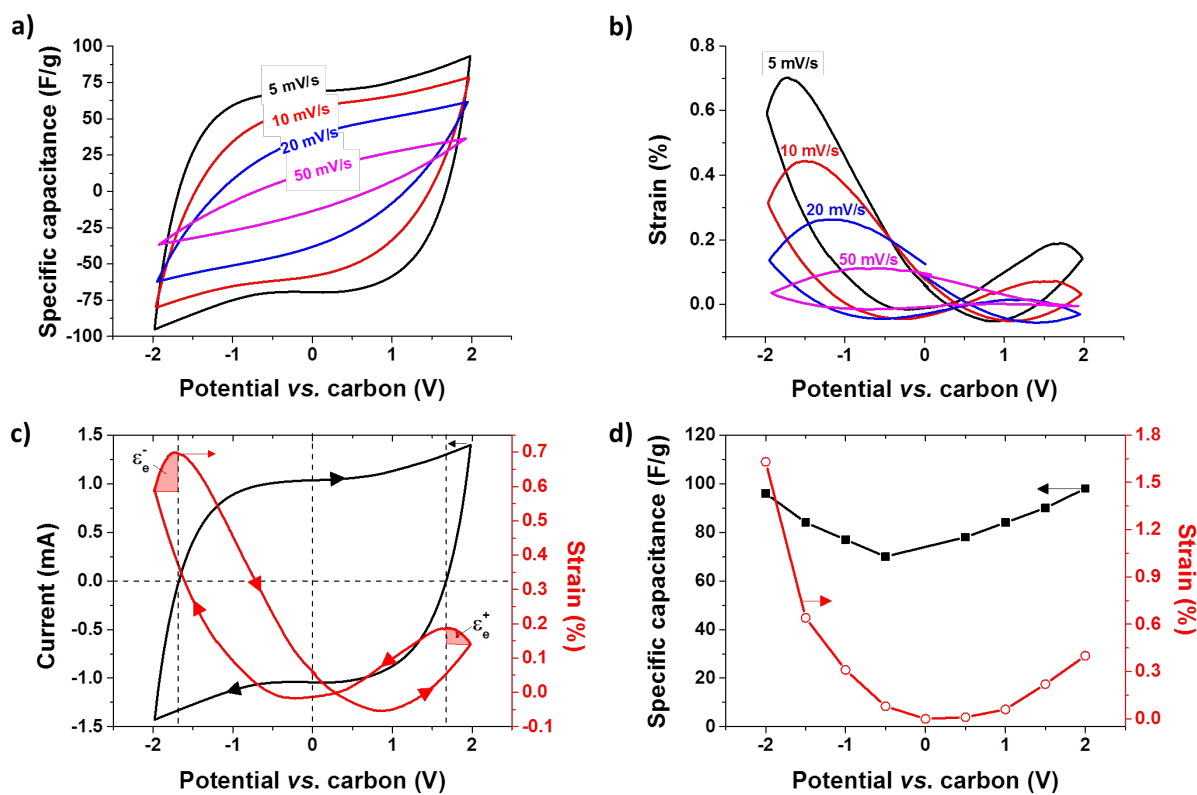

**Fig. 1:** Cyclic voltammetry data (**a-c**) and data from chronoamperometry (**d**). Cylic voltammograms, CV, (**a**) and strain *versus* potential curves (**b**) carried out at 5, 10, 20, and 50 mV/s in pure [Emim]$^+$ [BF$_4$]$^-$. Overlay of CV and expansion curve at a scan rate of 5 mV/s (**c**); the vertical and horizontal dashed lines represent zero current and/or zero potential. Specific capacitance and expansion derived from chronoamperometry (*i.e.*, charging at incremental potentials and discharging to 0 V for 1 h) (**d**).

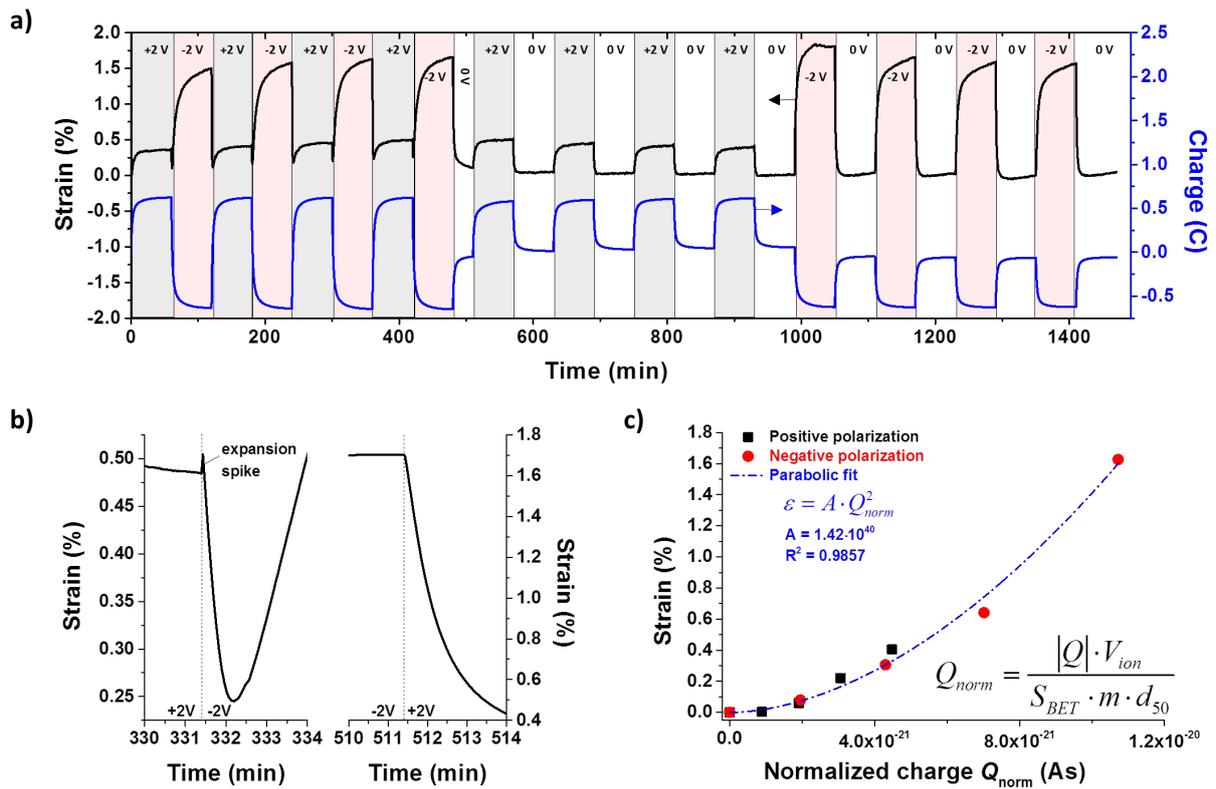

**Fig. 2:** Chronoamperometry data. Strain and accumulated charge during charging and discharging between -2 V and +2 V, -2 V and 0 V, and +2 V and 0 V (**a**). Magnification of the time of potential inversion and the corresponding expansion (**b**). Normalized charge *vs.* equilibrium strain and fitted quadratic function (**c**).